\documentclass[letterpaper]{sfchem}
\usepackage{graphicx}

\def\aap{{A\&A}}
\def\apj{{ApJ}}
\def\apjl{{ApJ}}

\newcommand\arcdeg{\mbox{$^\circ$}} 
\def\n2h{N$_2$H$^+$}
\def\c18o{C$^{18}$O}
\def\co{$^{12}$CO}
\def\13co{$^{13}$CO}
\def\h2{H$_2$}
\def\cc{cm$^{-3}$}
\def\cm2{cm$^{-2}$}
\def\kms{kms$^{-1}$}
\def\r{$\rho$}

\def\lc{\>\> ,}
\def\cz{C$^0$}

\begin{document}

\title{Large Scale Mapping of the $\rho$ Ophiuchi Region by SWAS}

\author{Di Li\inst{1} \and Paul F.\ Goldsmith\inst{2,3} \and Gary J.\ Melnick \inst{1}} 
  \institute{Harvard-Smithsonian Center for Astrophysics, Cambridge, MA, U.S.A.
  \and Astronomy Department, Cornell University
  Ithaca, NY, U.S.A.
 \and National Astronomy and Ionosphere Center,
  Ithaca, NY, U.S.A.} 
\authorrunning{Li, Goldsmith, and Melnick}
\titlerunning{Large Scale Mapping of the $\rho$ Ophiuchi Regions by SWAS}
 
\maketitle 

\begin{abstract}
	We have completed a 3 square degree CI $^3$P$_1$-$^3$P$_0$
map of the \r\ Ophiuchi regions using SWAS. The remarkably
stable receiver and backends systems of SWAS allow for uniformly calibrated
data set on this scale. Combined with \co, \13co, and \c18o\ maps
made using FCRAO, this data
set will facilitate a thorough study of the physics and the chemistry in this
nearby star forming region.
\keywords{Radio lines: ISM -- ISM: individual (\r\ Ophiuchi)}
\end{abstract}

\section{The Ophiuchi Region}
  At only 125 pc from the Earth (de Geus 1989), the \r\ Ophiuchi molecular cloud 
is one of the closest regions of active star formation.  This region has been
the focus of numerous 
observational investigations, as witnessed by mapping efforts by Wilking \& Lada (1983) and Loren et al. (1989),  to give only some references 
with extensive mapping data. At the same time, it was recognized to be an 
interesting area for studying stars at the earliest phases of their evolution, and also properties of interstellar dust (e.g.\ Vrba, Strom, \& Strom 1976).  
More recently, there has been a reawakening of interest, as witnessed by ISO observations in the
 infrared (Abergel et al.\ 1996; Liseau et al.\ 1999).

	Its overall size of approximately 15 pc  and its proximity present us a unique 
opportunity for studying both large scale structure and resolved 
dense cores. Our goal is to map the region in CI and three CO isotopologues (CO, \13co, \c18o).
 The optically thick  J=1-0 line of \co\ provides information on the kinetic temperature of clouds,
 and thus sheds light into the cloud thermal balance. The lines with modest opacity
such as  CI fine structure line
($^3$P$_1$-$^3$P$_0$)  and \13co\ J=1-0, are ideal for studying the large scale structure, particularly the 
effects of nearby ionizing sources, such as HD 147889. The largely optically thin tracer of \c18o 1-0 should provide a relatively complete census of cores from scale 
sizes of 1 pc to 0.03 pc. 

	We focus on the CI mapping of the \r\ Ophiuchi region in this discussion.
\section{Observations}
	The Submillimeter Astronomy Wave Satellite (SWAS) is particularly suitable to
carry out large scale mapping in CI. At 492 GHz, its 54$\times$68 cm antenna gives a 
relatively large beam of 3.5\arcmin$\times$5.0\arcmin\ full width half maximum (FWHM).
The off--axis Cassegrain design provides a 90$\%$ main beam efficiency, minimizing
the sidelobe pickup.
Although CI can be observed from ground based telescopes, 
the opacity and instability 
of the atmosphere at this frequency make it hard to perform position
 switching with large throws.
Finally, 
the acousto--optical spectrometer (AOS) on board SWAS has proven to be 
extremely stable. This is important for maintaining high quality of
 the overall calibration for a data set taken over a
time span of a couple of years.

	From February 1999 to March 2002, SWAS has taken data  toward
4345 individual positions
with 1.6\arcmin\ spacing. The Nyquist sampled maps cover a region of 3 square degrees
including the \r\ Oph A cloud and Lynds 1689. 
The data are convolved with a
Gaussian beam of 4.25\arcmin\ FWHM to mimic the true instrumental spatial resolution. The AOS channels 
have a nearly Gaussian response with FWHM channel width = 1.25 MHz, corresponding
to a velocity resolution of 0.76 \kms. The smoothed data have a 
representative RMS noise equal to 0.2 K per channel  
and most of the spectra are spectrally
resolved (Figure 1).
\begin{figure}[ht]
\resizebox{\hsize}{!}{\includegraphics{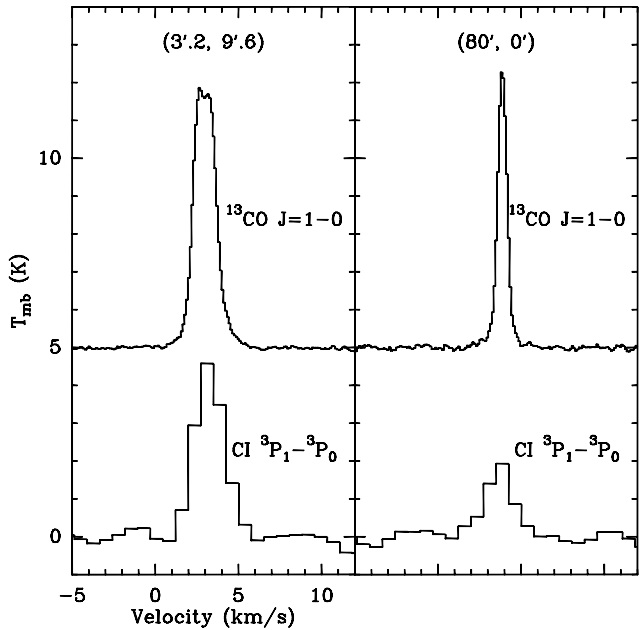}}
\caption{The CI and \13co spectra from the maps of the \r\ Ophiuchi clouds.
The map center is at (16$^h$26$^m$23$^s$.4, -24\arcdeg 23\arcmin 02\arcsec) (J2000).
Given in the top of each box is the angular offset ($\Delta RA$, $\Delta DEC$).
All spectra have been convolved with a Gaussian beam of 4.25\arcmin\ FWHM and
corrected for the main beam efficiencies of the respective
telescopes. The \13co\ line intensities have been reduced by a factor of 2 in order
to be shown on the same scale with CI. 
\label{fig1}}
\end{figure}

	Maps using the J=1--0 transitions of \co, \13co, and \c18o\ were made at the
Five College Radio Astronomy Observatory (FCRAO) during roughly
the same period as the SWAS observations. \co\ and \13co\ maps cover the same
3 square degree region. The \c18o\ data were only taken toward
selected areas with high column density. The details of these data will be
discussed elsewhere.

\section{Cloud Temperature and CI Column Density}

		For a spectral line with modest opacity like CI, the column density derived
 from the line integrated intensity can be parameterized as the following
\begin{equation}
 N_{CI}= N_1  F_\tau F_b F_u \lc
\label{n}
\end{equation}
where 
\begin{equation}
 N_1~(cm^{-2}) = 5.94\times10^{15} \int T_{mb} dV~ (km~s^{-1})
\label{n1}
\end{equation}
is the column density in the $^3$P$_1$ level (Frerking et al.\ 1989),

\begin{equation}
	F_\tau = \frac{\int \tau  dv}{\int(1-e^{-\tau}) dv}
\label{cftau}
\end{equation}
accounts for the line opacity,
\begin{equation}
	F_b = [1-  \frac{e^\frac{h\nu}{kT_x}-1}{e^\frac{h\nu}{kT_{bg}}-1}]^{-1} 
\label{cfb}
\end{equation}
accounts for the non-zero background temperature $T_b$, and
\begin{equation}
	F_u = \frac{1}{3}e^{23.6/T_x}+1+\frac{5}{3}e^{-38.8/T_x} 
\label{cfu}
\end{equation}
converts the population in the $^3$P$_1$ level to the total population.

 	Both $F_b$ and $F_u$ diverge for 
excitation temperatures $T_x < 5$ and remains essentially 
constant for $T_x>20$. For an average density of $n\sim10^{4}$ \cc\ and
and a nominal carbon abundance 
[\cz]/[CO]$\sim$0.1, the CI $^3$P$_1$-- $^3$P$_0$ line is 
close to being thermalized based on Large Velocity Gradient calculations. 
The opacity of this line 
can then be estimated using the antenna temperature and the gas temperature
under these conditions.
It is, therefore, important to determine the gas temperature.

	For cloud surfaces, the \co\ 1--0 antenna temperatures 
provides a measure of the gas temperature due to its large opacity. 
 The temperatures
thus obtained show a relatively smooth distribution in the range 
 20 to 25 Kelvins
 for the bulk part of the clouds. Patches of temperature enhancement (30--40 K) appear
around the center of \r\ Oph A and along some cloud edges of both \r\ Oph A and L1689.
A comprehensive explanation for the temperature structure should include both
internal and external heating. Particularly, the heating of the
cloud edges, if proven
exclusively due to external heating, could provide us a measure of the UV enhancement
(Li \& Goldsmith 2002). The change in the UV field must be tied with the gas
density and with the [\cz]/[CO] ratio in a consistent manner in developing a completed
physical and chemical model of the region.

	Overall, the physical conditions of \r\ Oph clouds
 produce thermalized CI emission at relatively
warm temperatures ($T_x > 20$ K), 
which restricts the variation in the combined factor $F_uF_b$ to be no 
more than 5$\%$.
One major uncertainty in the derived $N(CI)$ lies in the significant
opacity close to cloud centers, where the CI line may become optically thick and
the CO line shows self-absorption. For cloud edges,
the CO line may not be thick enough to allow an accurate
determination of gas temperatures. At the cloud edges, the density also
drops so that the CI line may no longer be thermalized.

	The derived CI column density shows a 
striking correlation in morphology
with that of the \13co (Figure 2). 
In low extinction regions, ring like structures around (120\arcmin, 0\arcmin)
and (-10\arcmin, -15\arcmin') are seen both in \13co\ and in CI. In high extinction
regions, CI and \13co\ peak at the same locations both in L1689 and \r\ Oph A.	
\begin{figure*}
\centering
\includegraphics[width=0.8\linewidth]{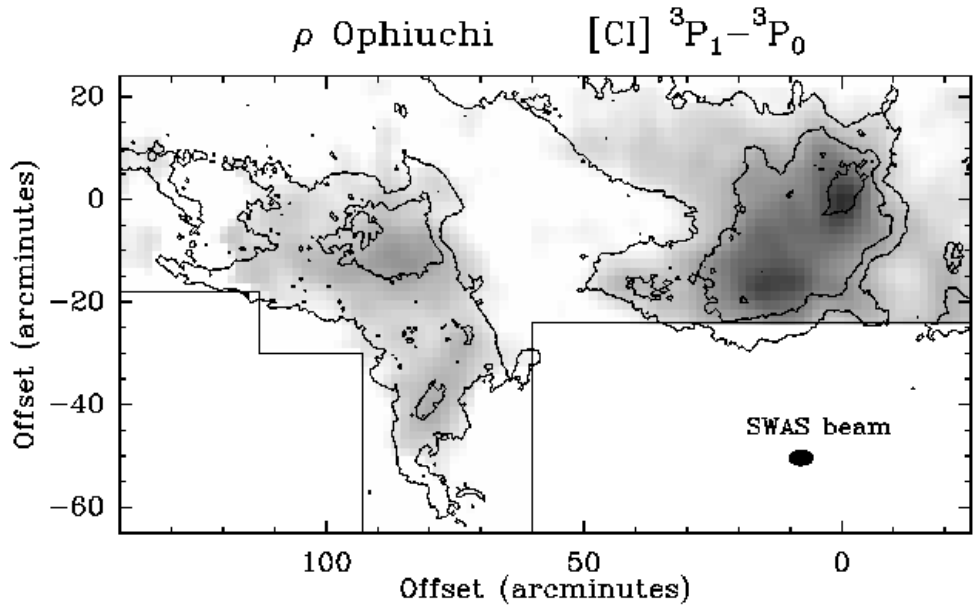}
\caption{Image of derived CI column density overlaid with the \13co\ column density.
The \13co\ contour levels are $9\times10^{15}$, $4\times10^{16}$, and 
$1.5\times10^{17}$ \cm2. The \r\ Oph A cloud is around (0\arcmin, 0\arcmin)
and L1689 is to the left. The peak CI column density is 
$2.8\times10^{17}$ \cm2.
\label{fig2}}
\end{figure*}
\section{Discussion}
	According to van Dishoeck and Black (1988), the fractional
abundance of CI 
increases from very low visual extinction ($A_v$) and to a peak at about
$A_v = 2$ for an average interstellar radiation field. CI disappears
in highly extincted regions before $A_v$ reaches 10. Apparently, a large
portion of the clouds are darker than $A_v=10$ with the  central extinction
of \r\ Oph A being a couple of hundred. If only a skin of CI is
seen toward the line of sights with large $A_v$, we would expect a
more uniform image of CI column density. Similar phenomena are 
also observed in other regions, particularly those with stronger UV
fields, such as M17 (Howe et al. 2000) and Orion (Plume et al.\ 2000).
To explain the presence of CI in high
extinction regions, suggestions have been made in terms of clumpy
structure and/or dynamics, which allow UV to penetrate further.
We will study the spatial correlation scales of CI and CO, which 
could test the plausibility of these two types of models

	As whole, CI and CO data form a valuable data set,
which allow us to pursue a consistent picture of the large
scale structure and chemistry of the \r\ Ophiuchi region.

\begin{acknowledgements}
This work was supported by NASA's SWAS contract NAS5-30702. The Five College Radio
Astronomy Observatory is supported by NSF grant AST 97-25951.
The National Astronomy and
Ionosphere Center is operated by Cornell University under a Cooperative
Agreement with the National Science Foundation.  
\end{acknowledgements}


\begin{thebibliography}{}
\bibitem[Abergel et al.(1996)]{abe96}
Abergel, A., et al. 1996, \aap, 315, L329
\bibitem[de Geus, de Zeeuw, \& Lub(1989)]{gzl89}
de Geus, E., de Zeeuw, P., \& Lub, J. 1989, \aap, 216, 44
\bibitem[Frerking, Keene, Blake, \& Phillips(1989)]{1989ApJ...344..311F} 
Frerking, M.~A., Keene, J., Blake, G.~A., \& Phillips, T.~G.\ 1989, \apj, 344, 311 
\bibitem[Howe et al.(2000)]{2000ApJ...539L.1337} Howe, J.~et al.\ 2000, 
\apjl, 539, L137 
\bibitem[Kenyon, Lada, \& Barsony(1998)]{klb98}
Kenyon, S.J., Lada, E.A., \& Barsony, M. 1998, \apj, 115, 252
\bibitem[]{}
Li, D., \& Goldsmith, P.F. 2002, submitted to ApJ
\bibitem[Liseau et al.(1999)]{lis99}
Liseau, R., et al. 1999, \aap, in press
\bibitem[Loren(1989)]{lor89} Loren, R.B. 1989, \apj, 338, 902
\bibitem[Motte, Andr\'{e}, \& Neri 1998]{man98}
Motte, F., Andr\'{e}, P., \& Neri, R. 1998, \aap, 336, 150
\bibitem[Plume et al.(2000)]{2000ApJ...539L.133P} Plume, R.~et al.\ 2000, 
\apjl, 539, L133 
\bibitem[Wilking \& Lada(1983)]{wl83}
Wilking, B.A. \& Lada, C.J. 1983, \apj, 274, 698
\bibitem[van Dishoeck \& Black(1988)]{1988ApJ...334..771V} van Dishoeck, E.~F.~, \& Black, J.~H.\ 1988, \apj, 334, 771 
\bibitem[Vrba, Strom \& Strom(1976)]{vss76}Vrba, F.J., Strom, S.E., \& Strom, K.M. 1976, AJ, 81, 317

\end{thebibliography}
\end{document}